\documentstyle[12pt]{article}

\begin{document}
\setlength{\unitlength}{0.25cm}
\thispagestyle{empty}

\hfill
\begin{tabular}{l}
JHU-TIPAC-920021\\
August, 1992
\end{tabular}

\vspace{0.4cm}

\begin{center}
{\large\bf
QUANTUM CORRECTIONS TO DEEP BAGS\footnote{
Supported by NSF grant PHY-90-96198
and by the Alfred P. Sloan Foundation.}}\\[1.0cm]

\setcounter{footnote}{1}
{\large
Stephen G. Naculich\footnote{Presented at the {\it XXVI International
Conference on High Energy Physics}, August, 1992.}
and Jonathan A. Bagger}

\vspace{0.5cm}

{\normalsize \sl
        Department of Physics and Astronomy\\
        The Johns Hopkins University\\
        Baltimore, MD 21218, USA}\\[0.6cm]

\end{center}

\vfill
\begin{center}
{\sc Abstract}
\end{center}

\begin{quotation}
Nontopological solitons, or ``bags,'' can arise
when fermions acquire their mass through a Yukawa
coupling to some scalar field.  Bags have played
an important role in models of baryons, nuclei,
and more recently, in the idea that a Higgs
condensate may form around a very heavy top quark.
It has been claimed that deep bags, which correspond to
tightly-bound states of fermions, will form
when the Yukawa coupling is strong.
Quantum corrections, however, are significant
in this regime.  We examine the effects of these
quantum corrections on the formation of nontopological solitons
in an exactly solvable large-$N$ model.  We find
that quantum bags differ dramatically from those
of the classical theory.  In particular, for
large Yukawa coupling, the bags remain shallow and
the fermions weakly bound.
\end{quotation}
\vfill

\newpage
\setcounter{page}{2}

\def\PLB{ \sl Phys. Lett. \bf B}
\def\NPA{ \sl Nucl. Phys. \bf A}
\def\NPB{ \sl Nucl. Phys. \bf B}
\def\PR{  \sl Phys. Rev.  \bf  }
\def\PRC{ \sl Phys. Rev.  \bf C}
\def\PRD{ \sl Phys. Rev.  \bf D}
\def\PRL{ \sl Phys. Rev. Lett. \bf}
\def\CMP{ \sl Commun. Math. Phys. \bf}
\def\Ann{ \sl Ann. of Phys. \bf}
\def\PRp{ \sl Phys. Rep.  \bf  }
\def\RMP{ \sl Rev. Mod. Phys. \bf}
\def\IJP{ \sl Int. J. of Mod. Phys. \bf }
\def\MPL{ \sl Mod. Phys. Lett. \bf}
\def\ZPC{ \sl Zeit. Phys. \bf C}
\def\NC{ \sl Nuovo Cim. \bf}
\def\LNC{ \sl Lett. Nuovo Cim. \bf}
\def\L{{\cal L}}
\def\eff{{\rm eff}}
\def\Ltwo{  \L^{(2)} }
\def\Lfour{ \L^{(4)} }
\def\bx { {\vec x} }
\def\eps{\epsilon}
\def\lam{\lambda}
\def\sig{\sigma}
\def\Gam{\Gamma}
\def\del{\partial}
\def\d { {\rm d} }
\def\e { {\rm e} }
\def\tr{ {\rm tr} }
\def\fer{ {\rm fermion} }
\def\sca{ {\rm scalar} }
\def\zet{ \zeta_a }
\def\epsa{ \epsilon_a }
\def\psii{ \psi^{i}}
\def\psiibar{{\overline \psi}^{i}}
\def\psibar{{\overline \psi}}
\def\gsim{\mathrel{\raise.3ex\hbox{$>$\kern-.75em\lower1ex\hbox{$\sim$}}}}
\def\lsim{\mathrel{\raise.3ex\hbox{$<$\kern-.75em\lower1ex\hbox{$\sim$}}}}
\def\cancel#1#2{\ooalign{$\hfil#1\mkern1mu/\hfil$\crcr$#1#2$}}
\def\slash#1{\mathpalette\cancel{#1}}
\def\pslash{ \slash{p}}
\def\delslash{ \slash{\partial} }

Nontopological solitons,$^{1-9}$
or ``bags,'' can arise
when fermions acquire their mass through a Yukawa
coupling to a scalar field.
Bags have been used to describe bound states of fermions.
In particle physics, the SLAC bag played an
early and important role in describing the confinement
of quarks inside hadrons.$^{3}$
In nuclear physics, nontopological bags have been
successfully used to model the binding of nucleons
within nuclei.$^{6,7}$
More recently, bags have also been discussed in conjunction
with the phenomenology of the heavy top-Higgs
system.$^{8}$

Nontopological~solitons are coherent
states in which the expectation value of
a scalar field is reduced
from its vacuum value by the presence of a fermion
field.  The solitons carry fermion number because
the fermion is energetically bound to the bag.
They are stable because they have lower energy than any
other configuration with the same quantum numbers.
The solitons form because the energy gained by
decreasing the fermion mass is greater than the
energy lost through the potential and gradient
terms in the scalar-field Hamiltonian.
At the classical level,
as the Yukawa coupling $g$ gets large, the
fermions become tightly bound inside deep bags,
whose energy and radius are independent of $g$.

In the full quantum theory, however, quantum
corrections can be very important,$^{1-9}$
especially in the nonperturbative regime of large $g$.
One must check to see whether bags still form.
There are two types of
quantum fluctuations to consider:  those in the scalar
field, which can destroy the coherent state, and those
in the fermion field, which can collapse the bag.

We examine bag formation in a consistent
quantum field theory.$^{9}$  We consider a theory with $N$ Dirac
fermions $\psii$ coupled to a real scalar field $\phi$.
We solve the quantum theory to leading order in the
large-$N$ expansion$^{10}$ for any value of the Yukawa coupling
$g$.  We find that the full quantum theory supports
nontopological bags.  The bags correspond to bound
states of $N$ fermions, with a binding energy of less
than about 5\%.  The quantum bags differ
significantly from those in the classical theory, where
the binding energy approaches 100\% for large $g$.  The
quantum corrections invalidate the classical picture of
tightly-bound fermions inside deep bags.

We first present our model.  The Lagrangian density is
\begin{eqnarray}
\L_0
&=& {1\over2}\, (\del_\mu \phi_0 )^2
- {\lam_0 \over 8N}\, (\phi_0^2 - u^2_0N)^2 \nonumber\\
& & + \sum_{i=1}^N \,\psiibar_0
    \left( i\delslash - {g_0 \over \sqrt{N} } \phi_0 \right)
    \psii_0 ,
\end{eqnarray}
where the subscripts signify bare quantities, and
all $N$-dependence is explicitly shown.  The
Lagrangian is characterized by three parameters,
$u_0$, $\lam_0$ and $g_0$.  The $N$-dependence is
chosen so that fermion-loop contributions are of
the same order as the tree-level couplings.  The
boson-loop contributions, however, are suppressed
by at least one factor of $N$.  This implies that
the bosonic fluctuations can be ignored, and the
scalar field can be treated as classical for any
value of the Yukawa coupling.

{}From eq. (1) we see that the field
$\phi_0$ develops a vacuum expectation value
$\langle\phi_0\rangle \ne 0$.  To compensate for
this, we shift $\phi_0 = \sqrt{N} v_0 + \sig_0$,
where $v_0$ is chosen so that $\langle \sigma_0\rangle=0$.
At tree level, $v_0$ is just $u_0$.
Then the mass of the field $\sigma_0$ is
$\mu_0 = \sqrt{ \lambda_0} v_0$,
while the fermion mass is $m_0 = g_0 v_0$.

To solve the model to leading order in $1/N$, we
compute all diagrams with a single fermion loop.$^{9}$
As usual, we must specify a renormalization
condition for each bare parameter.  We define the
wave function renormalizations in the standard way,
\begin{eqnarray}
{\d \Gamma^{(2)}_{\sigma\sigma}\over \d p^2}\bigg|_{p = 0} &=& 1 ,  \nonumber\\
{\d \Gamma^{(2)}_{\psi\psibar}\over \d\pslash}\bigg|_{p = 0} &=& 1 ,
\end{eqnarray}
where $\Gamma^{(2)}_{\sigma\sigma}$ and $\Gamma^{(2)}_{\psi\psibar}$
are the renormalized 1PI two-point functions for the renormalized
fields $\sigma$ and $\psi$.  We then fix the bare parameters
$u_0$, $\lambda_0$, $g_0$ and $v_0$ through the following
renormalization conditions:
\begin{eqnarray}
\Gamma^{(1)}_{\sigma} &=& 0, \nonumber\\
\Gamma^{(2)}_{\sigma\sigma}\big|_{p = 0}  &=&  - \ \mu^2, \nonumber\\
\Gamma^{(2)}_{\psi\psibar}\big|_{p = 0}   &=&  - \ m, \nonumber\\
\Gamma^{(3)}_{\sigma\psi\psibar}\big|_{p_i = 0}  &=&  -\
{g\over\sqrt{N}} .
\end{eqnarray}
The vanishing of the one-point function $\Gamma^{(1)}_\sigma$
ensures that we are expanding about the minimum of the
effective potential.  The other three
conditions define the renormalized masses of $\sigma$ and
$\psi$, as well as the renormalized Yukawa coupling $g$.

The Lagrangian (1) together with the renormalization
conditions (2) and (3) define the full quantum theory.
The quantum effective Lagrangian can be written as the sum of
two terms,
\begin{equation}
\L= \L_{\fer} + \L_{\sca} .
\end{equation}
The first term receives no quantum corrections,
\begin{equation}
\L_{\fer} = \sum_{i=1}^N \,\psiibar
    \left( i\delslash - {g \over \sqrt{N} } \phi \right)
    \psii\ ,
\end{equation}
where $\phi = \sqrt{N} v + \sigma$ and $v = m/g$.  The
second term is modified by the fermion loops.  It can be
written in a derivative expansion,
\begin{equation}
\L_{\sca}  = - V_{\eff}  +\Ltwo +\Lfour  +\cdots\ ,
\end{equation}
where the first term is minus the effective potential,
\begin{eqnarray}
          V_{\eff}
&=&  - \left( {\mu^2 \over 4} + {g^4 v^2 \over 8 \pi^2} \right)
          \left( \phi^2 - N v^2 \right) \nonumber\\
& & +   \left( {\mu^2\over 8 N v^2} + {3 g^4 \over 32 \pi^2 N} \right)
          \left( \phi^4 - N^2 v^4 \right) \nonumber\\
& &        - \ { g^4 \over 16\pi^2 N } \phi^4
         \ln \left( \phi^2 \over N v^2 \right),
\end{eqnarray}
and the two-derivative term is
\begin{equation}
\Ltwo =  {1\over2}
\left[ 1 - {g^2\over 8\pi^2} \,\ln \left (\phi^2 \over Nv^2 \right)
\right]    \left( \del_\mu \phi \right)^2
\end{equation}

As usual for Yukawa theories, the full quantum theory is
afflicted by many problems,$^{9}$ including a Landau
pole, a tachyon, and vacuum instability.  They result from the
fact that the theory is not asymptotically free, and
indicate that (4) must be viewed as the solution
to an {\it effective} theory, valid for energies and momenta
below some scale $\Lambda$.  The scale of $\Lambda$ depends
on the Yukawa coupling $g$.  For small $g$, $\Lambda$ is
exponentially large and can be safely ignored.  For large
$g$, however, $\Lambda$ plays an important role.  The size
of $\Lambda$ can be found by expanding the scalar field
propagator in powers of $p^2/m^2$,
\begin{equation}
\Gamma^{(2)}_{\sigma\sigma} = -\ \mu^2  +  p^2  +
{g^2 \over 80 \pi^2} {p^4 \over m^2}  +  \cdots
\end{equation}
For $\mu \ll m$, we find a tachyon pole at $ - p^2 \approx
80 \pi^2 v^2 \equiv \Lambda^2$.  Requiring $m \lsim \Lambda$ puts
a limit of $g \lsim 30$ on the Yukawa coupling.  Subject to
this condition, (4) describes the complete solution to
the quantum theory to leading order in $1/N$.

We will now demonstrate that the quantum theory supports
bag solutions.  In the large-$N$ limit, however, bags do not
form about a single fermion, but only when many fermions
are present.  We shall see that the bag lowers the
energy of the $N$-fermion state, indicating the
formation of a bound state, similar to a baryon in the
large-$N$ expansion of QCD.$^{11}$

We expect the lowest energy state of given fermion number
to have a static scalar field configuration.
The energy of the state is
\begin{equation}
E = E_{\sca} + E_{\fer} .
\end{equation}
For static configurations, the scalar energy is given by
\begin{equation}
E_{\sca} = -  \int \d^3 \bx \ \L_{\sca}\ .
\end{equation}
The fermion energy is found from the positive-energy solutions
to the Dirac equation in the presence of the scalar field $\phi$.
If $\zet$ denotes the spinor solution with (positive) energy
$\epsa$,
\begin{equation}
\left( -i{\vec\alpha}\cdot{\vec\nabla}\ +\ {g\over\sqrt N} \beta
\phi\right) \zet  =  \epsa \,\zet \ ,
\end{equation}
normalized so that
$\int \d^3 \bx\ \zet^{\dagger} \zet = 1$,
the total fermion energy is just
\begin{equation}
E_{\fer} = \sum_{a} \ n_a \,\epsa\ ,
\end{equation}
where the $n_a$ are the occupancy numbers of the
energy eigenstates, and $\sum_a n_a = N$ is the
total fermion number of the state.

In the lowest energy state with fermion number 1,
the scalar field has a constant expectation value,
$\langle\phi\rangle = \sqrt{N} v$,
and $\zeta$ is a free Dirac spinor of mass $m=gv$.
Thus, there is a solution to the quantum theory with
fermion number $N$ and energy $E = mN$,
consisting of $N$ free Dirac spinors
in a constant scalar field background.
For sufficiently large Yukawa coupling,
however, there are also soliton solutions,
with $\langle \phi\rangle < \sqrt{N} v$
in the presence of $N$ fermions.
The soliton is stable if this state has energy
$E < mN$.

For large $g$, there are two effects to consider:
the scalar field becomes strongly coupled to the
fermion, and the quantum corrections become important.
Therefore we proceed in two steps.  We first
examine the {\it classical} theory, which corresponds
to dropping the terms $\L^{(n)}$ with $n > 2$, as well
as dropping the terms in $V_\eff$ and $\Ltwo$
that depend on $g$.  In this case it is well-known
that the theory supports finite-energy bag-like solutions,
with $\langle\phi\rangle = \phi(r)$,
as shown in Fig.~1$a$.
(The solution is plotted
for the values $g = 25$ and $\mu = v$.  Our
numerical work was done with the aid of the program
COLSYS.$^{12}$)
The solution has fermion
number $N$ when the lowest orbital of each fermion is
occupied.
The energy per fermion is
$E/N = 6.3v \ll 25v$, so the $N$ fermions
are tightly bound.  Such configurations are known
as ``deep bags''
because $\phi(r)$ deviates significantly
from its vacuum value $v$.

We now consider the {\it quantum} theory.  For large
$g$, the quantum corrections significantly modify the
potential and the scalar gradient energy.
They also induce higher-derivative
terms in $\L_{\sca}$.  It is straightforward to solve
the quantum equations of motion.  We drop the terms
$\L^{(n)}$ for $n > 2$.
(We have checked that including $\Lfour$
changes the soliton energy by less than 0.25\% for
$g \lsim 30$  and $\mu = v$.)
As in the classical case, we find
finite-energy soliton solutions, as shown in Fig.~1$b$.
(The solution is plotted for $g = 25$ and
$\mu = v$.)  We see that the quantum corrections
dramatically alter the size and shape of the bag.
As above, the solution has fermion number $N$.  Now,
however, the energy
per fermion is $E/N = 23.8v$, so the fermions
are only weakly bound to the bag.

In Fig.~2 we see how the soliton energy scales with
$g$ for $\mu = v$.  For small $g$, the classical
and quantum bags are similar, with $E < mN$ for
$g\gsim 4$.  For larger $g$, the bags begin to
differ.  In the classical case, the energy is
independent of $g$ for large $g$.  In contrast,
the energy of the quantum bag scales as
$E/N \approx .95 gv$, while the radius goes as $R \sim
1/gv$.  The quantum corrections imply that the
fermions are weakly bound to a small and shallow
bag, with a binding energy approaching
about 5\% for large $g$.  In fact, a simple
scaling argument$^{9}$ shows that this
asymptotic 5\% binding energy is independent of
$\mu/v$.

In this model, the quantum corrections to the
energy have a simple physical origin which can
be understood in terms of the Dirac equation.
The presence of the bag changes the energy
eigenstates and eigenvalues.  It shifts
the valence orbitals {\it and} the Dirac sea
levels.
Equation (13) explicitly accounts
for the change in the valence orbitals.  The
shift in the Dirac sea is included implicitly,
through the quantum corrections to $\L_{\sca}$.
These corrections automatically sum
the shift in the Dirac sea.$^{2}$  To leading order
in $1/N$, the two effects give the entire
change in the energy.

The simplicity of the picture presented here
is a feature of the large-$N$
limit of the Yukawa theory.  For finite $N$, the bosonic
fluctuations modify the field equations for the
nontopological soliton solution.  This changes
the details of our picture, but it does not alter
our main point:  that quantum solitons can differ
dramatically from their classical counterparts.

We have used the large-$N$
expansion to find nontopological soliton
solutions to a quantum Yukawa theory.  For
large couplings, the energy of the quantum
bag scales with $g$.  This implies that bags
can indeed be used to model nuclei with their
relatively small binding energies.  In fact,
bag formation provides a powerful, nonperturbative
technique for finding bound-state solutions.
We have also found that quantum effects
deflate deep-bag solutions.  This raises
serious questions about using the SLAC bag as
a realistic picture of quark confinement.  If
we trust the general features of our results
all the way to $N=1$, we are also
led to conclude
that bag formation does not play a major role in
top quark physics.

\end{document}